# A position- and time-sensitive photon-counting detector with delay-line read-out

Ottmar Jagutzki[*a], Volker Dangendorf[b], Ronald Lauck[b], Achim Czasch[a], James Milnes[c]

[a]RoentDek GmbH, c/o IKF, Universität Frankfurt, Max-von-Laue-Str. 1, 60438 Frankfurt, FRG
[b]Physikalisch-Technische Bundesanstalt, Abteilung Neutronenphysik, 38116 Braunschweig, FRG
[c]Photek Ltd., 26 Castleham Road, St. Leonards on Sea, East Sussex, TN38 9NS, UK

**ABSTRACT**

We have developed image intensifier tubes with delay-anode read-out for time- and position-sensitive photon counting. The timing precision is better than 1 ns with 1000x1000 pixels position resolution and up to one megacounts/s processing rate. Large format detectors of 40 and 75 mm active diameter with internal helical-wire delay-line anodes have been produced and specified. A different type of 40 and 25 mm tubes with semi-conducting screen for image charge read-out allow for an economic and robust tube design and for placing the read-out anodes outside the sealed housing. Two types of external delay-line anodes, i.e. pick-up electrodes for the image charge, have been tested. We present tests of the detector and anode performance. Due to the low background this technique is well suited for applications with very low light intensity and especially if a precise time tagging for each photon is required. As an example we present the application of scintillator read-out in time-of-flight (TOF) neutron radiography. Further applications so far are Fluorescence Life-time Microscopy (FLIM) and Astronomy.

## 1. INTRODUCTION

Sealed image intensifier tubes with micro-channel plates (MCP) are widely used to intensify faint photon fluxes in the near-UV, visible and near-infrared wavelength regime [1]. Typically, first a photo-cathode converts photons into electrons. These photo-electrons are multiplied by the MCP and eventually reconverted to photons by a phosphor screen, preserving the lateral position of the initial photon impact on the photo-cathode. If such an image intensifier tube is equipped with double- or triple-stacked MCP operated in saturation, single photons can be imaged and counted.

There are similar photon detection devices, MCP-PMTs, where the photon-initiated electron cloud from the MCP stack is collected on a metal anode and an electrical signal can be picked up. A MCP-PMT can be considered as a classical photo-multiplier tube, but due to the use of a MCP stack as amplifying stage, a sub-nanosecond time resolution can be achieved over an extended area. Such devices are applied to count and determine the arrival time of single photons with picoseconds' precision, e.g. for time-of-flight measurements. A partitioned anode may even allow for a rather coarse position resolution [2]. On the other hand, the standard read-out technique for image intensifiers with high position resolution (phosphor screen with CCD read-out) does not allow for the simultaneous timing of individual photons with sub-nanoseconds' precision, as typically needed for most so-called *3d-imaging* applications (need for precise timing combined with two-dimensional imaging of single photons). Only if the photon rate is lower than the frame rate of the CCD chip (typically < 1000 Hz) it is possible to correlate individual photon positions with a timing signal picked up from the MCP stack and to allow for 3d-imaging.

Alternatively, one can pulse the photo-cathode or even the MCP voltage for achieving time information on a photon-emitting process. But this method can only select a single "time frame", which requires very intense light bursts to obtain complete images within a nanosecond long exposure time (e.g. spectroscopic neutron imaging from high power laser plasmas) or periodic processes with phase-synchronised exposure control of the gated intensifier [3]. Gated image intensifiers can be very effective under repetitive or triggered experimental conditions. However, outside the time window no photons are detected and multiple exposure time windows require either multiple independently gated intensified CCD cameras or consecutive measurements at different exposure time windows. Naturally, the number of cameras in a system is restricted due to cost and space reasons.

---

[*]jagutzki@atom.uni-frankfurt.de; phone +49 69 798-47001; fax +49 69 798-47107; www.atom.uni-frankfurt.de/jagutzki

At low photon fluxes the integrated systems with intensified CCD camera read-out are not suitable due to their rather low DQE (Detective Quantum Efficiency). The reason for this is the unavoidable read-out noise and dark current integration of the CCD. Here, in the field of single photon counting, the "ideal" single-photon detector should provide a fast read-out of individual photons, allocating to each detected event its 2-dimensional position coordinates and its arrival time (also called "3d-detection").

We will demonstrate that such devices are feasible with a position resolution of at least 1000x1000 pixels and sub-nanosecond time precision with negligible dead-time between individual photon hits. Furthermore, the photon detection (counting) rate capability should reach the physical limit given by the MCP stack's single-pulse operation mode (typical up to few MHz). Photon counting methods in principle do not show saturation effects at long exposure times. This yields better contrast and image dynamics than usually obtainable with intensified-CCD approaches, which makes photon counting imaging techniques superior even if no timing information is needed.

Applications for such photon counting detectors are for example found in fast-neutron radiography, which will be exemplarily discussed in more detail, Astronomy [4] (or similar low photon rate survey tasks) and in Laser-induced Fluorescence Life-time Imaging Microscopy (FLIM) [5] or spectroscopy (LIF). In the literature several approaches for achieving this goal by replacing the phosphor screen by a structured anode are discussed [5-10]. Like in MCP-PMTs, the charge cloud from the MCP stack is collected on a metal or resistive-layer anode without re-conversion to photons. In the simplest form, a pixel anode can be used with a separate timing electronic channel for each pixel. In order to reduce the number of electronic channels to a practically controllable amount, different anode types with only few specifically shaped electrodes or an anode with several corner contacts on a resistive coating can be used. The so-called resistive anode encoder (RAE) is widely used in some commercial systems [11,12].

A special so-called resistive screen (anode) with about 3 orders of magnitude higher area resistance than the RAE will be discussed in more detail here. It does not act as a position sensitive anode but produces a time-dependent image charge, which can be picked up by a dedicated structured electrode outside the image intensifier tube [4,13,14]. This facilitates the tube construction remarkably. However, it will be difficult to produce larger image formats than approximately 50 mm because the increasing thickness of the rear wall eventually leads to an unacceptable geometrical spread of the image charge signal.

All these approaches have in common that electrical signals from the distinct anode contacts are picked up and their relative pulse heights or time sequence for each detected photon event allow for the localisation of the charge cloud centre position. In a proximity-focussed tube with an MCP-based electron amplifier this is in good approximation the position of the photon impact on the photo-cathode. These "position signals" can be easily correlated with a fast time signal picked up from the MCP stack, which is related to the impact time of the photon on the photo-cathode with sub-nanosecond precision.

So far, none of these attempts could satisfactorily meet all design goals of the "ideal" (3d) single-photon detector as defined above. We consider the delay-line technique to be the most promising read-out concept because very large format detectors (e.g. 75 mm or even 150 mm active diameter) providing high position resolution can be built. Furthermore, this method potentially allows reaching higher photon counting rates than read-out techniques based on relative charge measurement and even correlated photon hits within few nanoseconds can be analysed by a special triple-layer delay-line anode [15].

Delay-line anodes of different designs have been used in combination with image intensifier tubes by other groups [8-10]. For the single-photon detectors with 75 mm and 40 mm active area and internal delay-line anode presented in this paper we have chosen the helical-wire delay-anode design, which has been proven reliable for open-face MCP detectors and which are especially suited for large delay-line arrays (e.g. large detector formats) [16,17]. For the resistive-screen image intensifier, we have designed economical read-out electrodes on standard PCB material.

In this paper we investigate the performance of image intensifier tubes equipped with an internal helical-wire delay-line anode array and, alternatively, tubes with a resistive screen and an external pick-up electrode with delay-line read-out. Furthermore, we show the performance of such image intensifier tubes for fast-neutron radiography, emphasising the importance of photon timing information for such a "pure" imaging task.

We conclude with an outlook on the perspectives of these detector types in becoming the "ideal" single-photon detector in near future.

## 2. THE MCP PHOTO-MULTIPLIER TUBE WITH HELICAL-WIRE DELAY-LINE ANODE

Fig. 1 shows a sketch of the PMT with internal helical-wire delay-line anode (DL-PMT). The helical-wire delay-line (HWDL) anode is placed a few millimetres behind the MCP. Connectors for the wire terminals are guided through the rear wall of the sealed tube. The remaining components, the MCP stack and the photo-cathode, are fabricated, positioned and contacted like in any typical MCP-PMT assembly. Once the technique of producing sealed MCP-PMTs of large and variable format is established, it is rather straightforward to implement a custom anode behind the MCP stack. But even if most care is taken during the production process, the risks for having a leakage in the ceramics/glass/metal housing assembly will increase with tube size and with the number of electrical feed-through pins. Proper choice of material and a robust design is very critical so that the baking and scrubbing processes during tube production do not affect the mechanical and electrical properties of the anode and the quality of the vacuum is well preserved.

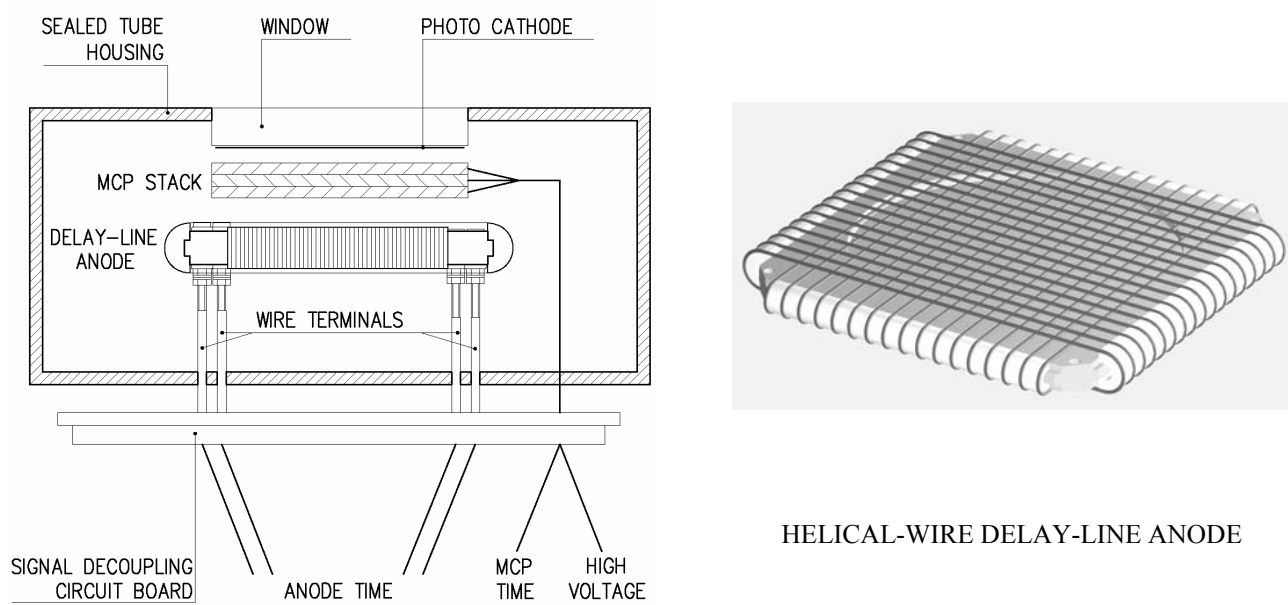

HELICAL-WIRE DELAY-LINE ANODE

Fig. 1: Sketch of the PMT with helical-wire delay-line anode (DL-PMT).

Two DL-PMT prototypes of different sizes, entrance window materials and photo-cathodes have been built and tested. The DL80-PMT is equipped with a low-noise S-20 photo-cathode of 75 mm diameter on fibre optic window. For the DL40-PMT a red-enhanced S-20 photo-cathode on fused silica window was chosen. The MCP stacks with respective sizes consist of three MCP with 10 micron pores and 60:1 L/D (channel length/diameter ratio). Due to this choice it was possible to achieve the high gain and excellent single-photon pulse-height distribution (see fig. 2). This is an important property of this kind of detectors: The pulses of every photo-electron must be sufficiently high for allowing a proper discrimination against electronic noise by selecting an appropriate electronic threshold level. It should be noted that in spite of the high MCP stack quality only about 75% of the emitted photo-electrons are registered because not all of them give rise to an electron avalanche in the MCP stack. Thus, the total detection efficiency is reduced to approximately 75% of the photo-cathode's quantum efficiency.

The HWDL, the read-out electronics and the position encoding method is described extensively in an earlier publication [16]. In short, wire pairs are wound around a support plate in form of a double-helix. Two perpendicular oriented layers allow for a two-dimensional position read-out. The electron charge cloud is collected on the positively biased wire of each pair, almost equally shared between the two layers. The induced electrical signals in the wires (each wire pair forms a differential delay-line) are transmitted to the terminals and are decoupled by capacitors located outside the tube. The four differential signals from the wire array ends are transformed into single-ended 50 Ohm impedance signals by adequate circuits.

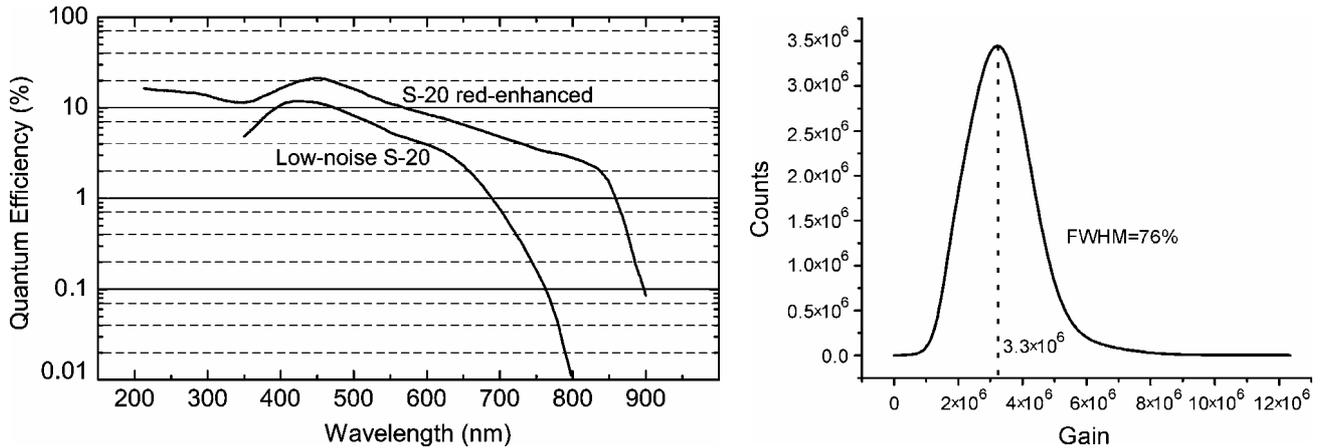

Fig. 2: Pulse-height distribution (gain) of the DL40-PMT's MCP stack at 3000 V bias (right) and photon conversion quantum efficiencies (QE) of the chosen photo-cathodes/window materials as function of the photon wavelength (left). The decrease at short wavelengths for the DL80-PMT (low-noise S-20) is caused by the vanishing transparency of the fibre optic window for UV-light. The total detection efficiency is about 75% of the photo-cathodes' quantum efficiencies.

The delay between the signals on both ends of each delay-line is proportional to the position of the charge cloud centre's projection in the respective dimension. The position resolution depends on the time precision for determining this delay. In order to measure the relative signal delays with high precision, these signals of about 10 mV mean pulse height and few nanoseconds' width are amplified with fast low-noise amplifiers. Standard Constant Fraction Discriminator (CFD) circuits are used to discriminate against electronic noise and to determine the relative signal times with 50 ps precision or better.

A fifth, position-independent, timing signal can be obtained from the MCP stack: Due to the momentary drain of charge during the electron avalanche emission, a fast voltage drop with positive polarity (few 10 mV signal height on 50 Ω termination with 1-2 ns width) occurs and is picked up via a capacitor from the MCP stack's rear side. This signal is likewise amplified and electronically processed.

The time sequence of these five signals contains the information on the lateral photon absorption point (in the photo-cathode) and the detection time. This time sequence is registered by a multi-channel time-to-digital converter (TDC) and the position and time information for each photon detection event is computed by a CPU for real-time visualization and event-by-event storage.

The DL40 and DL80 HWDL anodes inside the tubes were adopted from standard "open" MCP-delay-line detector systems as used for the detection of charged particles in a vacuum environment. The tests focussed not so much on a detailed detector characterisation but on a feasibility study: Can this technique be extended to sealed photo-tubes maintaining the performance characteristics known from the "open" detectors?

With the photo-cathode kept at ground potential, a voltage dividing resistor chain provides all necessary voltages for the MCP stack and the wire array from a single high-voltage supply via an adequate shielded cable. The voltage-dividing and signal-decoupling circuits are placed just outside the sealed tube (covered by resin for safety) and deliver the timing signals via coaxial cables to amplifier/CFD circuits of type *ATR19* and TDCs of types *HM1-B* and *TDC8HP* [17], which are designed for MCP/HWDL read-out of "open" detectors.

The images presented in figures 3 and 4 are 2d-histograms from a PC-memory. They were accumulated by measuring each photon's signal time sequence, calculating the time difference for both dimensions in the PC RAM and incrementing the corresponding cell of the histogram for every photon counted. The bin size is given by the TDC's channel width of 133 ps (*HM1-B*) or 25 ps (*TDC8HP*). This corresponds to position bins from the delay/mm scaling function of the respective delay-line: about 0.75 ns/mm for the DL40 and about 1 ns/mm for the DL80. However in the following, the "raw" images without position calibration are presented.

Fig. 3 shows images obtained with the DL40-PMT: A shadow mask of 1 mm holes every 3.8 mm was placed in contact with the window and homogenously illuminated (left). The "dark emission" of thermally emitted electrons from the red-enhanced S-20 photo-cathode (about 13000 counts/cm²·s at room temperature) is shown on the right. The DL80-PMT's low-noise S-20 photo-cathode produces < 30 counts/cm²·s dark emission. Both values are within expectations.

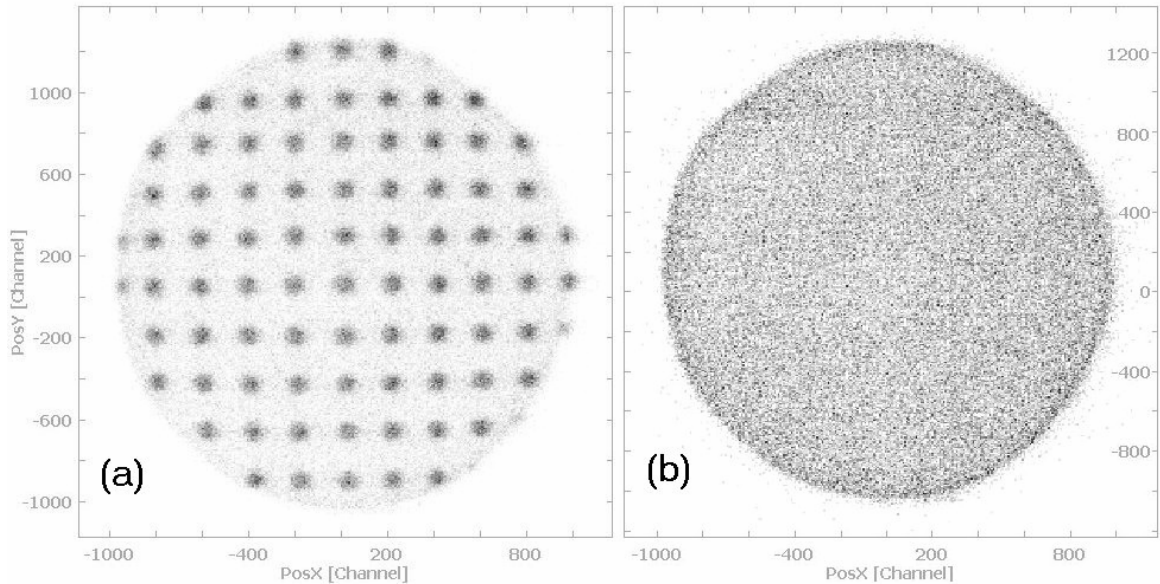

Fig. 3: Images (linear gray scale) of the DL40-PMT: shadow of a homogeneously illuminated mask of 1 mm holes every 3.8 mm, placed in contact with the entrance window (a), and dark noise image (b).

It is to note that a PMT with standard S-20 and other photo-cathodes with large sensitivity in the red should be cooled to about -20$^o$C in order to reduce the dark count rate to a tolerable level for photon counting. This is a common practise, although it was not applied during the tests presented in this paper.

Fig. 4 shows the image response of the DL80-PMT: A shadow masks with 1 mm wide lines every 5 mm (left) demonstrates the homogeneous and uniform imaging properties as were expected from tests on similar "open" delay-line detectors. Minor non-linearity can arise from an imperfect wire array and from electrostatic lens effects near the outermost image area.

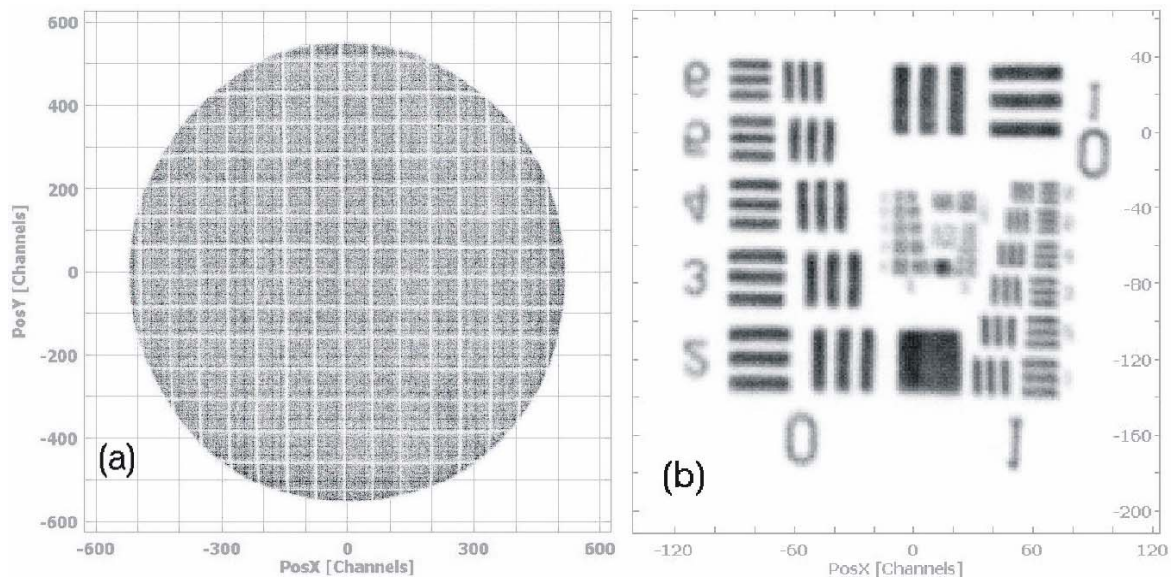

Fig. 4: Image response of the DL80-PMT to illuminated shadow masks: grid lines of 1 mm width every 5 mm (a), NASA test mask (b). The field-of-view in (b) is approximately 20x20 mm, both images in linear gray scale.

The position resolution of the DL80-PMT in combination with the *HM1-B* TDC could be determined with a standard NASA mask (right) to a FWHM of about 150 micron. This is in accordance with expectations for the effective pixel width of the *HM1-B* (75 micron with DL80). It corresponds to a RMS or pixel resolution of about 1000x1000. By the choice of a "better" TDC (like the *TDC8HP*) the resolution can be further increased. However, this could not yet be demonstrated during the tests on the DL40-PMT with *TDC8HP* due to the high thermal background (the *TDC8HP* was not yet available during tests of the DL80-PMT).

It is expected from results on open delay-line detectors that 1000x1000 pixels resolution for the DL40-PMT and 2000x2000 pixels for the DL80-PMT is achieved in combination with the *TDC8HP* or a similar highly resolving TDC with LSB < 50 ps. The photo-electron rates during all these tests where kept between 1000 counts/s and 1 million counts/s because higher count rates could not be registered by the TDCs. No saturation effects or image degradation are observed in this regime. Tests on the absolute time resolution could not be done due to the lack of a pulsed photon source. However, since the position is determined through timing signals, one can expect that the good imaging characteristic coincides with an equally well timing response, as known from the "open" MCP detectors with HWDL.

## 3. THE RESISTIVE-SCREEN IMAGE INTENSIFIER WITH DELAY-LINE READ-OUT

Image intensifiers with optical read-out are an industrial standard product and can be produced very economically in specialised facilities. Introducing major changes to the tube design (like introducing a dedicated read-out board into the high vacuum tube) usually requires major modification to the production facility and makes these tubes quite expensive. Furthermore, once built and sealed, there is no chance to modify or optimize the read-out electrode unless one starts producing a new tube.

Therefore, in 1999, we have proposed and patented a method to minimize the changes required to a commercial image intensifier and simultaneously permitting a very flexible design of the read-out electrode [18]. This method, originally proposed by Battistoni *et al.* [19] and extensively used in various applications with gaseous detectors, uses an anode (screen) with a very high surface resistance, which replaces the usual metal-coated phosphor screen of a standard optically read out image intensifier. This resistive layer with a typical surface resistance of > 1 M$\Omega_\square$, acts as an electromagnetically transparent window for the signal of the travelling electron cloud as it propagates from the last MCP to the resistive screen. While in gaseous detectors these layers are made of resistive lacquer based on carbon and plastic material, the high purity requirements in vacuum image intensifiers demand for non-outgassing materials like Germanium. Fig. 5 shows a sketch of such a Resistive-Screen Photo-Multiplier Tube (RS-PMT). Apart from the resistive screen and a possible thinner and not necessarily optically transparent output window it is identical to a commercial image intensifier with a double or triple MCP stack.

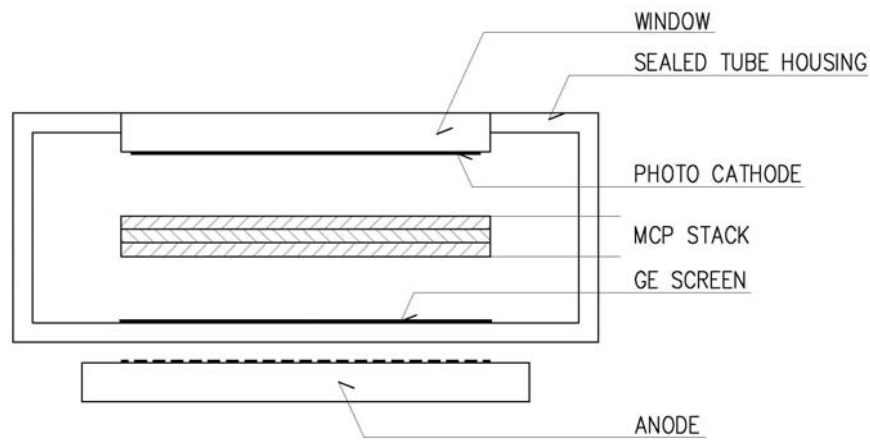

Fig. 5: RS-PMT tube with resistive screen (here: Germanium layer) replacing the phosphor screen of a conventional image intensifier. The position information is obtained by a dedicated pickup electrode (anode) outside the tube housing.

The resistive layer of Germanium is deposited directly on the output window (glass or ceramics) of the intensifier. The read-out board is mounted outside the tube in close contact to this window. The spacing between the travelling charge cloud inside the tube and the separated read-out electrode outside causes a geometrical spread of the induced signal on the read-out board. For MCP-based image intensifiers but also for gas detectors like GEMs or RPCs (resistive plate chambers) this is of great benefit because it allows using rather coarse read-out structures, e.g. strips with few millimetre pitch, for a delay-line read-out. In recent years various read-out techniques using this resistive screen technique for image intensifiers were reported in literature [4,14,18].

Also several theoretical studies to calculate the induced charge distribution are available in literature [20-22]. We have experimentally studied and optimised the relevant parameters, including the geometrical distribution of the induced signals and its dependence from the distance between resistive layer and pickup electrode and the required surface resistance of the layer for maximum signal transmission. Furthermore, we studied the geometrical structure of the signal pickup strips of the read-out board. Detailed results of these measurements will be published elsewhere [23]. Here we show only one example, which is of some importance for the further development of RS-PMTs.

Fig. 6 shows a one-dimensional projection of the radial distribution of the induced charge behind a 1 mm thick output window and a 1 mm deep induction gap, i.e. the space between MCP or GEM and the resistive anode screen. The FWHM here is 2 mm, which is more or less exactly the value one needs for a read-out board with a strip pitch of 2 mm. Larger distances will increase the width of this distribution, which will result in a smaller amplitude of the induced charge and cause dispersion of the signal rise-time at the end of the delay lines. Both factors will result in a deterioration of the position resolution. Unfortunately, the output windows of image intensifiers have to be of a certain thickness due to the requirement of a stable housing for the vacuum inside the tube, which is usually above the optimal value. Nevertheless, future developments must address this issue and an optimisation of the material and its thickness is unavoidable especially for proposed larger devices than the presently planned tubes with 40 to 50 mm sensitive diameter.

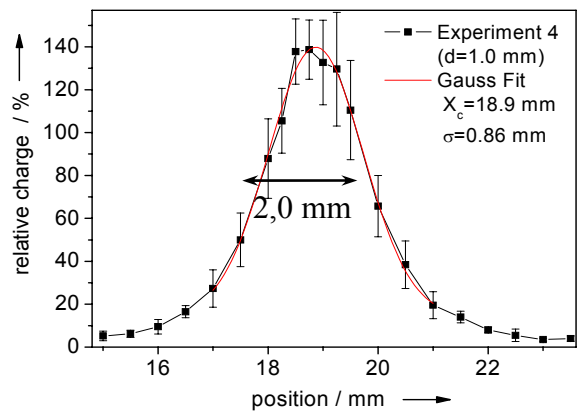

Fig. 6: Radial width of the induced charge signal at 1 mm distance between resistive screen and read-out electrode.

In our tests we have used a RS-PMT with 25 mm active diameter, produced by Proxitronic GmbH in Bensheim, Germany [1]. The position read-out electrode is a double-layer standard printed circuit board (PCB) with orthogonal strip structures on both sides. The principle structure of this electrode was proposed earlier by Eland [24] and was applied for RS-PMT read-out by Jagutzki *et al.* [13]. Here we present some modifications to improve its handling and performance. To ensure equal charge sharing and minimum capacitive coupling between both sides of the read-out board we have optimized the width and the shape of the strips.

Fig. 7 shows the front- and backside structure of the optimized read-out electrode. The strips, whose shapes resemble a chain of diamonds, are copper-printed on a 0.5 mm thick FR4 board. The strips have a pitch of 2 mm. The front side (Fig. 7a) has much smaller pads to avoid electromagnetic shielding of the back side (b) and to ensure equal signal amplitudes on both sides of the read-out board. The diamonds of the front side are located in the open areas between the diamonds of the back side. The structures overlap only at the intersections of the thin strips connecting the diamonds, minimising capacitive coupling between both sides of the board. The strips on each side are connected via an LC delay-line of 2.5 ns delay per tap.

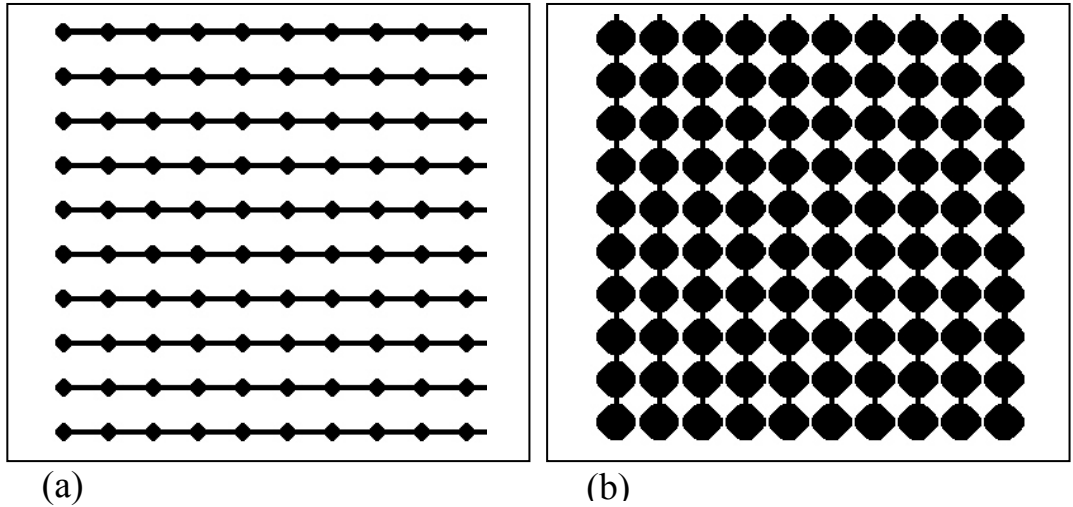

(a)                          (b)

Fig. 7: sections of the strip structure of the front (a) and back (b) side of the position read-out board. The strips have a pitch of 2 mm and run in orthogonal directions.

As an example of results obtained with this tube and anode fig. 8 shows an image of a printed optical shadow mask, illuminated by a pulsed hydrogen lamp (a) and the projection of an image of a single pinhole with same illumination (b). The position resolution obtained with this device and described read-out is 87 micron (FWHM). The resolution on the back-side of the board is slightly worse (116 μm FWHM). It must be noted that the width of the original pinhole is only of minor influence to this distribution because in each light pulse from the lamp many photons are converted in the photo-cathode. The delay-line method measures the centre of charge, therefore, what we see here, is basically the intrinsic resolution of the RS-PMT and the read-out method.

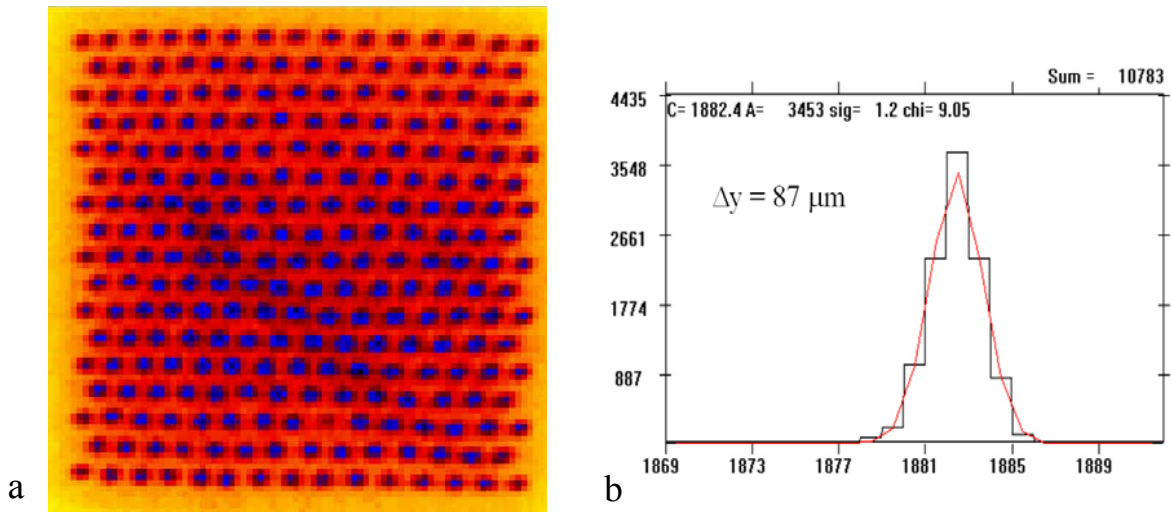

Fig. 8: Image of an optical mask measured with the RS-PMT and delay-line read-out (a). The mask on the RS-PMT is illuminated by a pulsed $H_2$-lamp. The distance between the holes of the mask is 0.7 mm, the holes' diameter is 0.3 mm. b) shows the projection of a pinhole. The width of the peak corresponds to 87 μm (FWHM). The resolution on the back side of the board is slightly worse (116 μm FWHM).

More elaborate tests of this RS-PMT can be found elsewhere [4,25]. In the following chapter we present a practical application of a RS-PMT in pulsed fast-neutron imaging for explosives and drug detection.

# 4. FAST-NEUTRON RADIOGRAPHY WITH POSITION- AND TIME-SENSITIVE PHOTON DETECTORS

Energy-resolved fast-neutron radiography is a promising method for element-sensitive imaging of low-Z materials, which cannot be resolved by standard X-ray or thermal-neutron techniques. Fast neutrons have a high penetration power and, due to the specific structures (resonances) in the neutron cross sections of elements like C, N or O in the MeV range, the spatial distribution of those elements, e.g. in air cargo containers or luggage, may be resolved by this method.

A prerequisite for the application of the method is the possibility of obtaining neutron transmission images in selected narrow energy ranges. In recent years we have investigated various methods of spectroscopic 2d-imaging in pulsed, broad-energy neutron beam illumination by applying time-of-flight (TOF) methods for energy selection [26-28]. Here we present some results with the above-described RS-PMT and read-out electrode, which is used to image the light emission pattern of a scintillating neutron converter screen with high time resolution (few ns) for each detected neutron.

Fig. 9 shows a schematic view of the detector. The neutron converter is a 10 mm thick BC400 scintillator slab. Out of the 20 x 20 cm$^2$ large scintillator an optical system projects a circular area of about 10 cm in diameter to the image sensor. The optical read-out consists of a large aperture lens with focal length $f$ = 120 mm, focal/diameter ratio 0.95 (F-number) and the RS-PMT (same as used in the tests before) as image sensor.

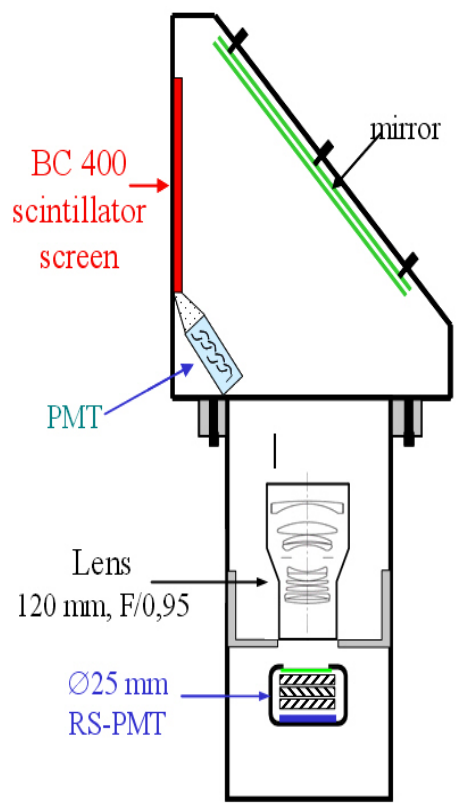

Fig. 9: Schematic view of the neutron imaging system with an RS-PMT.

With this detector we have performed measurements at the neutron beam facility of PTB, similar to the ones described in [26,27]. The neutron beam was produced by a pulsed 12 MeV deuterium beam that strikes a 3 mm thick Beryllium target. The distance between neutron source (Be target) and detector was 6 m and the sample was placed directly on the input window in front of the scintillator.

Fig. 10 shows the neutron images of a sample consisting of carbon rods (a) of various diameters (∅) and lengths (*l*) plus a steel wrench. For the radiographic image (b) neutrons in the whole available continuous energy range (1 -10 MeV) were utilized. The carbon rods and the steel wrench are visible in (b). In order to produce the element-sensitive image (c), the coordinates and the TOF of each neutron are stored in a list-mode file. In the off-line analysis, images for defined TOF windows are collected, which correspond to the large structures in the energy-dependent neutron cross section (for details see [28]). After appropriate digital image processing one can produce images like 10c, which only resemble the distribution of a certain chemical element (here: carbon). Note that the steel wrench, visible in 10b, has disappeared from the image.

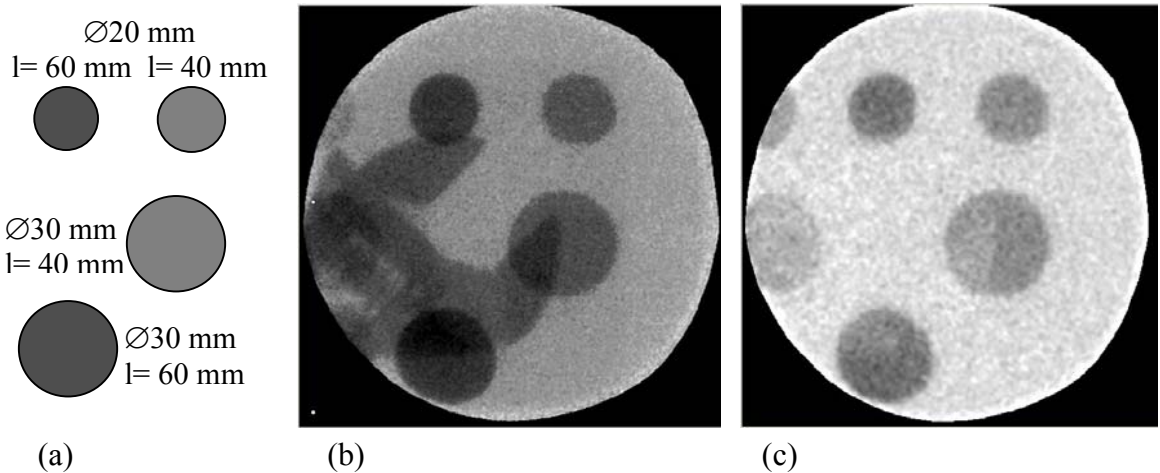

Fig.10: Neutron images of various carbon rods and a steel wrench. (a) shows the configuration (length *l* and diameter ∅ of the carbon rods). (b) shows the image in a broad-energy neutron beam (1 – 10 MeV) and (c) the carbon distribution after utilizing the resonance imaging technique.

A complete neutron imaging system would consist of a matrix with many elements of such scintillator/lens/RS-PMT assemblies for achieving a sufficient large area, which is required for this application [27]. Here, the use of the RS-PMT would be of great advantage compared to the DL-PMT described above due to its potential of economical mass production. The comparably inferior position resolution is not of importance in this application because the neutron interaction process and the range of the secondary charged particles limit the available resolution.

## 5. SUMMARY AND OUTLOOK

In summary, the results of these tests confirm that all performance characteristics of open delay-line detectors are reproduced in DL-PMT-type photon detectors and that known properties from standard image intensifier tubes (photo-cathode response, gain) are also preserved for a DL-PMT. It is thus possible to design DL-PMT tubes for a specific application (photo-cathode type, effective detection diameter, etc.) and scale the typical performance parameters from experiences on existing detectors. It should be straightforward to increase the effective image diameter to at least 120 mm and to implement a *Hexanode* delay-line type for multi-hit applications [15].

Next-generation TDC circuits will allow increasing the detection rate to more than 1 megacounts/s. Thus it can be expected that the "ideal" time- and position-sensitive single-photon detector can soon be realized using DL-PMT-type image intensifiers. Due to the superior imaging properties of pulse counting methods it would be desirable to also use such detectors for pure imaging applications. This seems feasible for photon rates at least up to 10 megacounts/s.

If the demand for the active area is below 50 mm and if large quantities of tubes are required at reasonable costs, the resistive-screen image intensifier tube can be an alternative. The external delay-line anode shows the same performance characteristics as the internal helical wire delay-line anode. The robust tube design will allow a qualification for harsh environment and the imaging performance is not influenced by strong magnetic fields.

Not yet finally determined is a possible rate limitation of the RS-PMT due to charge build-up on the resistive screen. This detector version has already demonstrated its potential in several applications and is ready to be used in different photonic fields with similar time and position detection tasks or for pure single photon counting and imaging applications.

# ACKNOWLEDGEMENT

The Authors would like to thank Jürgen Barnstedt from the Astronomy Institute of the University in Tübingen, FRG for the provision of an RS-PMT.